\documentclass[manuscript]{aastex}

\usepackage[varg]{txfonts}
\usepackage{graphicx}
\usepackage[]{natbib}
\shorttitle{Anti-correlated Soft Lags in GX 339-4}
\shortauthors{Sriram et al.}

\begin{document}
\title{Anti-correlated Soft Lags in the Intermediate State of Black Hole Source GX 339-4}
%\subtitle{I. Proof for truncated disk}

\author{K. Sriram\altaffilmark{1}, A.R. Rao\altaffilmark{2} and C. S. Choi\altaffilmark{1}}
%\affil{Korea Astronomy and Space Science Institute, Deajeon 305-348, South Korea}
%\author{}

%\email{astrosriram@yahoo.co.in}
%\author{A.R. Rao\altaffilmark{2}}
%          
%\affil{Tata Institute of Fundamental Research, Mumbai 400005, India}

%\author{C. S. Choi\altaffilmark{1}}
%\affil{Korea Astronomy and Space Science Institute, Deajeon 305-348, South Korea}
%\author{}
  \altaffiltext{1}{Korea Astronomy and Space Science Institute, Deajeon 305-348, South Korea. astrosriram@yahoo.co.in}
\altaffiltext{2}{Tata Institute of Fundamental Research, Mumbai 400005, India}

\begin{abstract}
{We report the few hundred second anti-correlated soft lags between soft and hard energy bands in the source
GX 339−4 using RXTE observations. In one observation, anti-correlated soft lags were observed using the
ISGRI/INTEGRAL hard energy band and the PCA/RXTE soft energy band light curves. The lags were observed
when the source was in hard and soft intermediate states, i.e., in a steep power-law state.We found that the temporal
and spectral properties were changed during the lag timescale. The anti-correlated soft lags are associated with
spectral variability during which the geometry of the accretion disk is changed. The observed temporal and spectral
variations are explained using the framework of truncated disk geometry. We found that during the lag timescale,
the centroid frequency of quasi-periodic oscillation is decreased, the soft flux is decreased along with an increase
in the hard flux, and the power-law index steepens together with a decrease in the disk normalization parameter.
We argue that these changes could be explained if we assume that the hot corona condenses and forms a disk in the
inner region of the accretion disk. The overall spectral and temporal changes support the truncated geometry of the
accretion disk in the steep power-law state or in the intermediate state.} 
\end{abstract}

\section{Introduction}

The huge database of RXTE observations of different Galactic
black holes (GBHs) shows that the spectral and timing properties
of the accretion disk in these objects can be constrained
depending on the classification schemes (Esin et al. 1997;
Grove et al. 1998; McClintock \& Remillard 2004; Belloni
2010; McClintock et al. 2009). During the rise and fall of the
outburst the accretion disk is found to be in the low–hard (LH)
state where the hard component (power-law index Γ ∼ 1.5)
is strong, most probably due to the presence of a Compton
cloud or a jet, along with an extremely weak or undetected soft
component (presumably due to a Keplerian disk). The power
density spectrum (PDS) in this state shows a strong band-limited
noise, occasionally with quasi-periodic oscillations (QPOs)
at low frequencies. On the other hand, when the soft X-ray
intensity increases, the hard component becomes very weak
and the soft component dominates the spectrum along with
the disappearance of a QPO feature in the PDS. This state
is known as the thermal-dominated (TD) state (Kubota et al.
2001; Remillard \& McClintock 2006) or the high–soft (HS)
state (Belloni 2010). Both classification schemes point toward
a similar picture, where the accretion disk approaches close to
the black hole, when the source moves from the LH to the TD
state.
The evolution of the accretion disk from the LH state to
the HS state and vice versa is not only complicated but also
poorly understood. It is generally believed that when the mass
accretion rate increases from the LH state, the strength of the
soft component increases relative to that of the hard component,
which softens. Since the spectrum is steep (power-law index
Γ $\ge$ 2.5) and, in most of the GBHs, the X-ray luminosity is
high, this state is termed the steep power-law (SPL) state or
the very high (VH) state (Miyamoto et al. 1991; Remillard \&
McClintock 2006). The complexity of this state is unveiled by
study of the hardness intensity diagram (HID), which is found to
be a useful phenomenological tool to trace the evolution of the
accretion disk during an outburst (Belloni et al. 1996; Mendez
\& van der Klis 1997; Belloni 2010). This unified scenario is
further constrained by radio observations (Corbel et al. 2000;
Gallo et al. 2003; Fender et al. 2004; Fender et al. 2009).
This transitional state is further divided into two states, i.e.,
a hard intermediate state (HIMS) and a soft intermediate state
(SIMS); for further details, see Belloni 2010 and references
therein. X-ray and radio studies show that the HIMS–SIMS
transition can be approximated as being due to the crossing
over of a “jet line” in the HID; however, further observations
in this state are needed to constrain this picture as the track of
the jet line in the HID is still not clearly understood (Fender
et al. 2009). In the temporal domain, the most important property
of this state is the presence of high-frequency QPOs $\ge$
100 Hz (HFQPO) together with the intermediate (IM) QPOs in
the frequency range of 1–10 Hz (Casella et al. 2004; Remillard
\& McClintock 2006; Belloni 2010). In the spectral domain,
perhaps the most important physical property of the SPL/IM
state is the presence of a low temperature ($\le$ 10 keV), high optical
depth (2--5) Compton component, possibly coming from a
compact cloud close to the black hole (Done \& Kubota 2006;
Sriram et al. 2007, 2009; Caballero-Garcıa et al. 2009). This
is found that as the disk approaches or recedes from the black
hole, the QPO frequency increases or decreases (Done et al.
2007). The broadband spectrum shows a hard component which
is possibly due to thermal Comptonization of soft photons in
the Compton cloud (Done \& Kubota 2006; Sriram et al. 2007,
2009). This could be the most favorable mechanism in this
state because both the soft and hard components dominate and
this spectral characteristic is less prominent in other spectral
states. In this state, the temporal and spectral transitions are
fast and distinct (these states cover a large fraction of the HID
plane) when compared to the other states (Dunn et al. 2010).
It is found that in the IM/SPL state the accretion disk is truncated
and hence the coupled soft/hard emission regions show 
anti-correlated lags which constrain and strengthen the truncated
geometry model.
There have been numerous studies on lags that range from
a few milliseconds to a few hundred seconds. Small-scale (few
milliseconds) positive and negative lags have been found in GRS
1915+105 (Cui 1999; Lin et al. 2000; Reig et al. 2000). Previous
researchers have found that the observed lags are frequency
dependent. Similar results were obtained for XTE J1550−564
(Cui et al. 2000). More detailed studies were carried out forXTE
J1859+226, which shows that B-type QPOs always show a hard
lag, whereas the A-type and C-type QPOs show only the soft
lag. They argued that the energy dependence of QPOs rules out
the possibility of a simple disk origin and that the 6 Hz QPO is
related to unknown fundamental processes (Casella et al. 2004).
Recently, it was found that, in the case of GRS 1915+105, the
lags are both frequency and energy dependent (Qu et al. 2010),
which rules out the existing possible models of low-frequency
QPO production. These small Fourier lags strongly favor the
Comptonization process close to the black hole and the lag time
corresponds to the energy gain time in the Compton cloud or
probably in the base of the jet.
The few hundred second scale lags were first reported in
the source Cyg X-3. The cross-correlation between the soft
and hard energy bands shows anti-correlated hard lags along
with spectral pivoting (Choudhury \& Rao 2004). Later, similar
anti-correlated hard lags were found in several black hole
sources, GRS 1915+105, XTE J1550−564, and H1743−322
(Choudhury et al. 2005; Sriram et al. 2007, 2009), and in one
neutron star source, Cyg X-2 (Lei et al. 2008). It was found
that along with spectral pivoting the centroid QPO frequency
is shifted during the observed lag timescale. The soft and hard
fluxes were anti-correlated during these lags and altogether the
results suggest that, as the disk front moves toward the black
hole, the soft photons cool the Compton cloud, changing the
geometrical or physical size. Until now, soft lags were often
observed in smaller dynamical scales but anti-correlated soft
lags were not observed on a longer timescale.
Here we report the discovery of anti-correlated soft lags in the
black hole source GX 339−4. GX 339−4, amicroquasar, has an
orbital period of $\sim$1.7 days, with a mass function f(M) = 5.8 $\pm$ 0.5 M$_{\odot}$ 
(Hynes et al. 2003). The secondary companion is an Ftype
subgiant star situated at a distance of 6--15 kpc (Hynes et al.
2004; Zdziarski et al. 2004). Based on spectroscopic studies,
the binary inclination of this system was found to be low, i.e.,
i = 15$^{o}$ (Wu et al. 2001). GX 339−4 may harbor a black hole
with a high spin parameter of {\it a} = 0.8--0.9 (Miller 2007). The
radio observations during the state transition suggest discrete
ejection events along with relativistic radio jets with speed
v/c $>$ 0.9 (Gallo et al. 2004). GX 339−4 was the first source to
exhibit features of the VH state (Miyamoto \& Kitamoto 1991).
The source had an outburst during the year 2006–2007 which
was continuously monitored by the RXTE satellite. The outburst
lasted about 10 months during which RXTE made 220 pointed
observations. The detailed spectral and temporal analysis of 83
observations covering the transition from the LS state to the HS
state has been carried out by Motta et al. (2009). They found
that the high-energy cutoff initially decreased during the hard
IM state and increased in the soft IM state, implying a sudden
change in the high-energy radiative mechanism close to the
black hole. This outburst was also observed by the the XMMNewton
and International Gamma-Ray Astrophysics Laboratory
(INTEGRAL) satellites and the results are presented in Del Santo
et al. (2009) and Caballero-Garcıa et al. (2009).
We have used HIMS/SIMS classified observations to find
the lags between the soft and hard energy bands using RXTE
observations. In three observations, we found anti-correlated
soft lags of the order of a few hundred seconds along with a
shift in the centroid frequency of the QPO. We also found that
the spectral properties changed during these observations. In one
of the observations where anti-correlated soft lag was found, the
PCA/RXTE soft energy band (2--5 keV) and ISGRI/INTEGRAL
hard energy band (20--40 keV) also simultaneously showed anticorrelated
soft lag. The temporal and spectral variations during
the detected lags are explained in the framework of a truncated
disk geometry. Re-condensation of hot flow is proposed to
understand the shift in the QPO frequency and spectral changes
in the respective observations.

\section{Data Reduction}

The 2006–2007 outburst in GX 339−4 was extensively
observed by the RXTE satellite (Swank 1999). The satellite is
comprised of three onboard detectors, the Proportional Counter
Array (PCA; Jahoda et al. 2006), High-Energy X-ray Timing
Experiment (HEXTE; Rothschild et al. 1995), and All-Sky
Monitor (ASM; Levine et al. 1996). Detailed temporal and
spectral studies have been carried out during the outburst and
the respective observations were classified into three canonical
states (Motta et al. 2009). Motta et al. found that out of 83
observations, 15 belong to the HIMS and 4 belong to the SIMS
classification. We have obtained the cross-correlation function
(CCF) for the HIMS and SIMS classified light curves. We
have used the PCA data, which show the characteristics of the
HIMS and SIMS classification scheme. For spectral studies,
we have extracted the spectra from the top layer of the PCU2
data, which is the most well calibrated among the PCUs, and
0.5\% systematic errors are added to the PCA spectrum. To
study the PDS, we used the single-bit data mode. We have
used HEASOFT v6.8 software to reduce the data and XSPEC
v12.5 (Arnaud 1996) for the spectral fittings.
During this outburst, the INTEGRAL satellite observed the
source on five different days (for more details, see Caballero-
Garc´ıa et al. 2009). INTEGRAL (Winkler et al. 2003) carries on
board two main gamma-ray instruments, the Spectrometer on
INTEGRAL (SPI, 20 keV–8 MeV; Vedrenne et al. 2003) and the
Imager on-Board INTEGRAL Satellite (IBIS, 15 keV--10 MeV;
Ubertini et al. 2003; Lebrun et al. 2003), as well as two
monitoring instruments in the X-ray and optical range, the Joint
European X-RayMonitor (JEM-X, 3--35 keV; Lund et al. 2003)
and the Optical Monitoring Camera (OMC; Mas-Hesse et al.
2003). We have used INTEGRAL/ISGRI data (SCW ObsID:
053600360010) for further reduction and analysis.We have used
the IBIS\_SCIENCE\_ANALYSIS main script to obtain the light
curve in the 20--40 keV energy band. Data reduction was carried
out using the Off-line Science Analysis (OSA) v8.0 package.
To obtain the lag time we closely follow the method described
in Sriram et al. (2007). An inverted Gaussian function is used
to fit the cross-correlation spectra and the error bars shown are
at the 90\% confidence level throughout this paper.

\section{Analysis}
\subsection{Temporal Analysis}
The background-subtracted light curves were obtained in two
energy bands, the soft (2--5 keV) and hard (20--50 keV) energy
bands. We have used the crosscor program to look for the anticorrelated
lags and discovered anti-correlated soft lags of a few hundred seconds in the three observations (Figure 1). In
one case, the cross-correlation is performed between the PCA/
RXTE soft energy band and the nearly simultaneous ISGRI/
INTEGRAL hard energy band (20--40 keV). The CCF is found
to be asymmetric and an anti-correlated soft lag is observed
(Figure 2).
For the three observations shown in Figure 1, the derived values
of anti-correlation (minimum values of CCF) are -0.44 $\pm$ 0.11, -0.47 $\pm$ 0.12, and -0.35 $\pm$ 0.10, respectively. The probabilities
of getting more than these values, due to random fluctuations
in the data, are 1.09 $\times$ 10$^{-4}$, 3.73 $\times$ 10$^{-5}$, and 1.60
$\times$ 10$^{-3}$, respectively (Bevington \& Robinson 1992). The derived
values of the lags are -95 $\pm$ 40 s, -1068 $\pm$ 62 s, and
-858 $\pm$ 108 s, respectively. Treating the estimated error as 1σ
(it should be noted that we have used the criterion of $\Delta \chi^{2}$=4.0, typically signifying a 90\% confidence level), the lags are
detected at confidence levels of 2.4$\sigma$, 17.2$\sigma$, and 7.8$\sigma$, respectively,
for the three observations. The corresponding value for
the INTEGRAL/PCA anti-correlation has a CCF value of -0.49
$\pm$ 0.21 (significant at a level of 5.90$\times$10$^{-2}$), with a lag detected
at -352 $\pm$ 102 s (a 3.5 $\sigma$ detection). Hence, we can conclude
that the lags are detected at a very high significance level.

It is quite possible that black hole sources like GX 339--4
have long time variabilities, and sampling these light curves
at shorter timescales (a few thousand seconds of observations)
might cause spurious peaks in the CCF when a “high” point
in one light curve hits a “high” or “low” point in another light
curve. But the fact that the anti-correlations are observed in the
RXTE as well as the INTEGRAL data shows that the observed
anti-correlations are not due to any instrumental artifacts. Further,
the low- and high-energy light curves pertain to the same
source and hence they are unlikely to show totally independent
variations. In order to rule out such spurious possibilities,
we have generated two independent simulated light curves of
long duration (220 s), using the method of Timmer \& Koenig
(1995). The simulated light curves were produced using the
mean (1415 counts s$^{-1}$), standard deviation (15 counts $^{-1}$),
and PSD power-law index ($\beta$ = -1.30) of the soft energy band
light curve of the second observation (Figure 1). The two light
curves are resampled as the original data (bin size of 64 s). A
randomly varying number with mean of 1.0 (standard deviation
0.20) was generated using the “randomn” function available in
Interactive Data Language (IDL).We smoothed this series using
the “SMOOTH” function (available in IDL). The “SMOOTH”
function filters an input series using a running-mean top with
a given kernel size. We have used kernel size 3 to smooth the
series. After doing this the standard deviation was found to be
0.12. Then, one of the light curves is multiplied with the obtained
smoothly varying number (variation is random in nature)
with a mean of 1.0, which is uncorrelated to the real variability.
We have taken 3000 s segments from the simulated light curves
and perform cross-correlation analysis. We found that most of
the CCFs show anti-correlation and positive-correlation coefficients$\sim$$\pm$
0.25 and a few segments of the CCFs show low anti
($\le$--0.4) and high positive ($\ge$0.4) cross-correlation coefficients
(Figure 3). The observed (Figure 1) anti-correlation coefficient for the first and second observations is around $\le$-0.4 and the
simulation result shows that only 6 segments (out of 350 segments;
only 1.8\%) show anti-correlation coefficients less than
$\sim$ -0.4. This indicates that the observed anti-correlation is significant
at $>$98\% confidence level. Apart from the “SMOOTH”
function, we have used the “MEDIAN” function and the SAVGOL
(Savitzky--Golay) filter (both available in IDL) to smooth
the varying number of mean 1. The SAVGOL filter uses a
weighted moving average technique. Using both the techniques,
we carried out the simulation and found that 8–10 segments (out
of 350 segments; 2.2\% and 2.8\%) show an anti-correlation coefficient
less than $\sim$-0.4.

     \begin{figure}
\centering
%\resizebox{\textwidth}{1}   
\includegraphics[height=15cm,width=12cm,angle=-90]{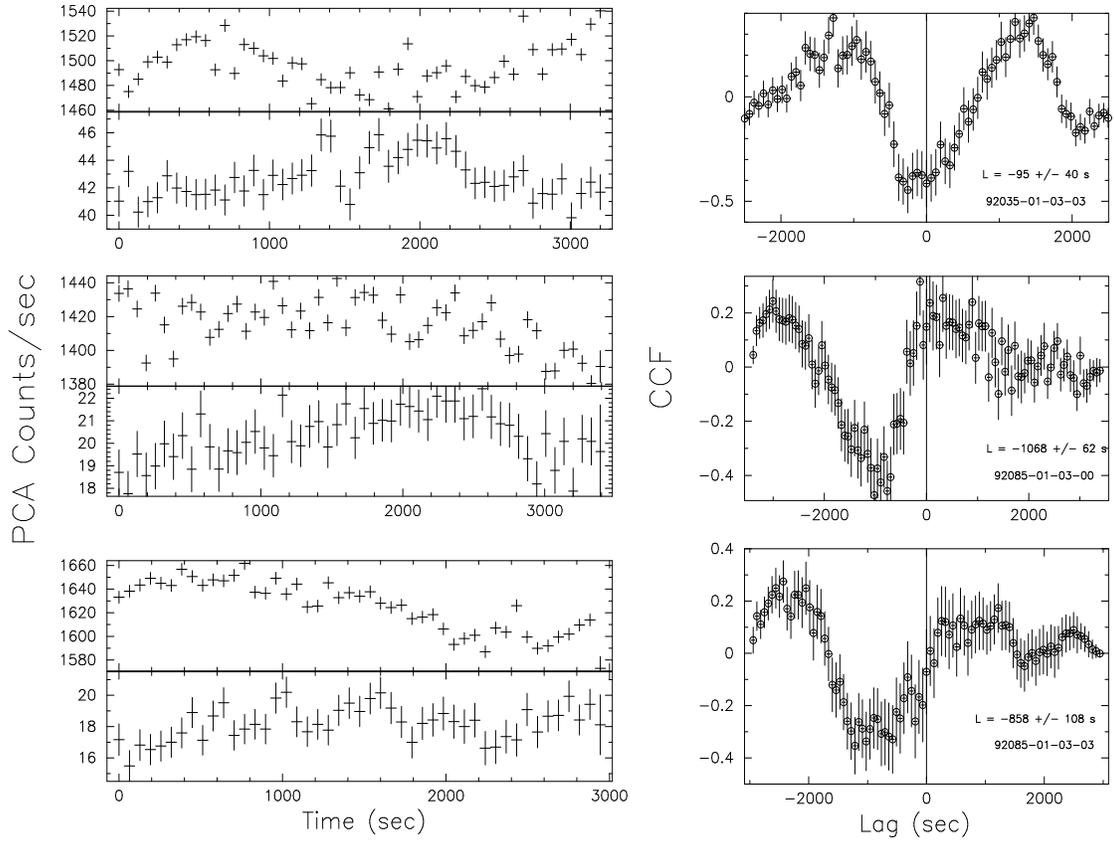}
     \caption{Left: The PCA soft (2 -- 5 keV, top) and hard (20 -- 50 keV,
 bottom) X-ray light curves of GX 339-4 are shown for three observations.
Right: Cross-correlation functions are  plotted for the respective observations
and the vertical line represents zero lag.  The measured lags (L) and
observation Id for each observations are also shown.} 
       \label{Fig1}
 \end{figure}

\begin{figure}
\centering
%\resizebox{\textwidth}{1}   
\includegraphics[height=15cm,width=12cm,angle=-90]{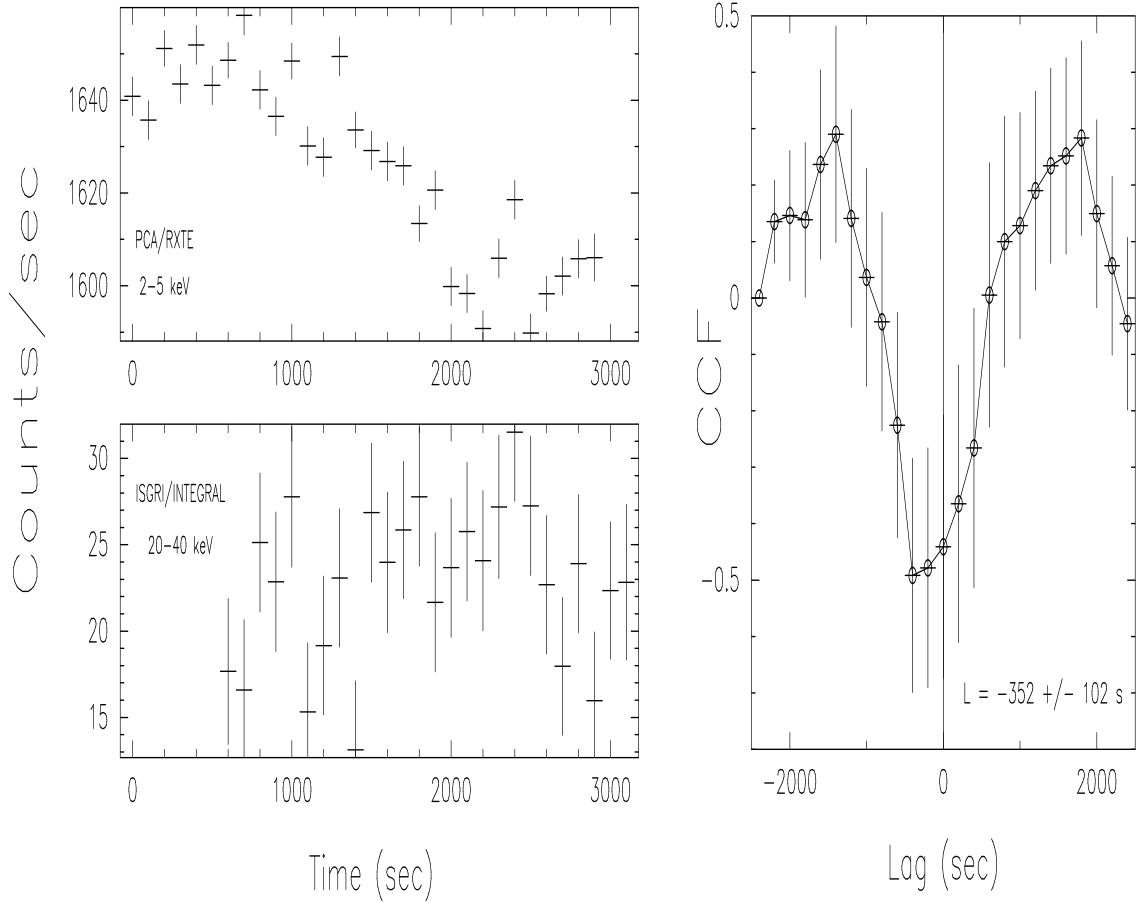}
     \caption{Left: The PCA/RXTE soft (2 -- 5 keV, top, ObsId: 92085-01-03-03) and ISGRI/INTEGRAL hard (20 -- 40 keV, bottom, SCW ObsId: 053600360010) 
X-ray light curves are shown. Right: The cross-correlation function is plotted. 
The vertical line represents zero lag and 'L' is the lag time scale.} 
       \label{Fig2}
 \end{figure}

\begin{figure}
\centering
%\resizebox{\textwidth}{1}   
\includegraphics[height=15cm,width=12cm,angle=-90]{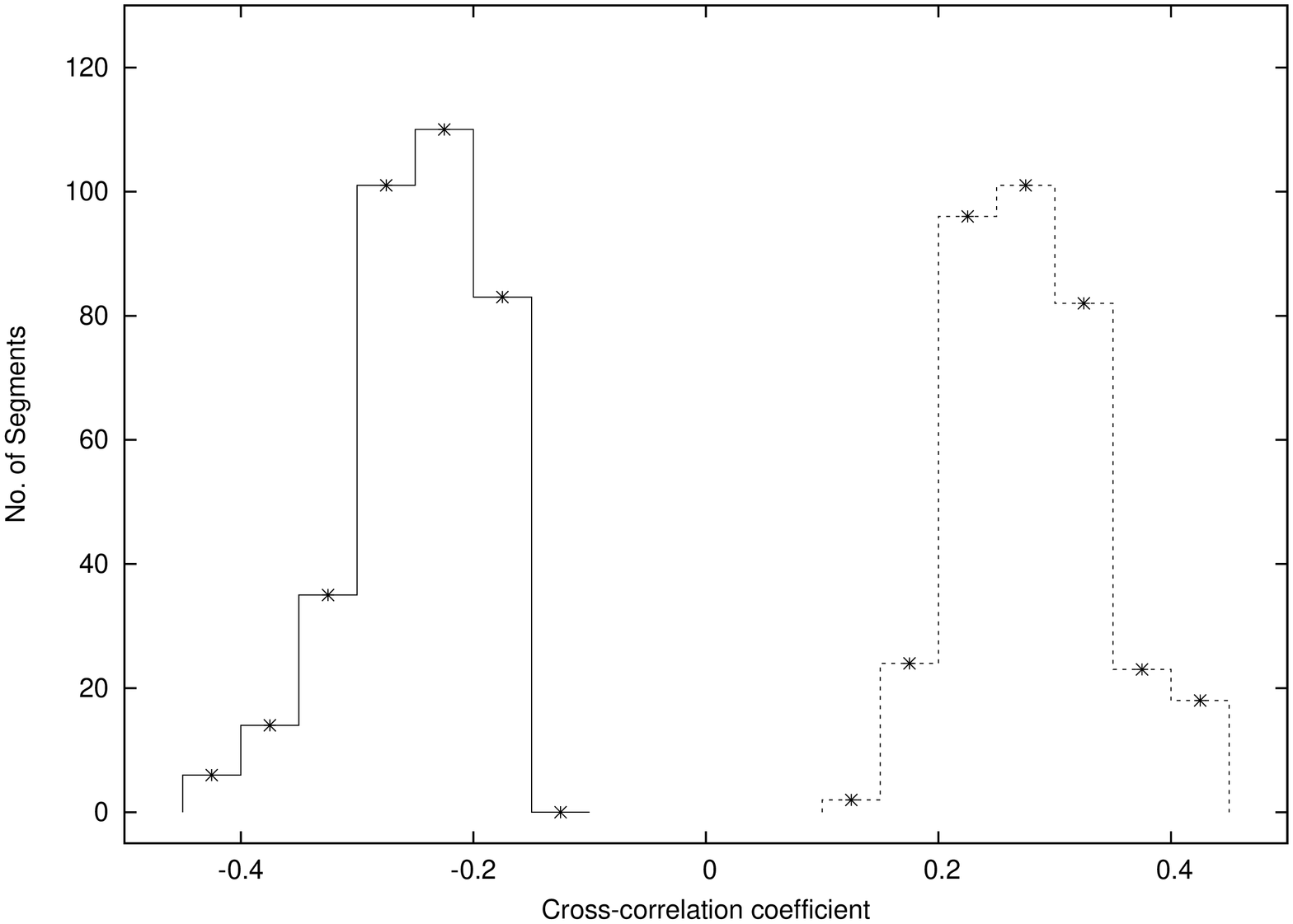}
     \caption{The histogram show the anti correlation (thick line) and positive correlation (dashed line) coefficients vs number of segments, each of 3000 s. 
The number of segments with an anti-correlation coefficient $\le$-0.4 occupies minute fraction 
of area in the histogram and are only 1.8\% of total segments which indicates that the observed anti-correlation are not spurious.}
%only 30\% among 
%segments which show strong correlations and only 9\% including segments whose CCF show low degree of variability (uncorrelated). } 
       \label{Fig3}
 \end{figure}
GX 339−4 is the first black hole source to show anticorrelated 
soft lags at such large timescales. The soft lag means
that the soft photons are lagging compared to the hard photons.
The observation time (MJD 54163) of the third observation
(ObsID: 92085-01-03-03) matches the peak of a small magnitude
hard re-flare as depicted from both HEXTE and Swift/BAT
light curves (Motta et al. 2009; Caballero-Garc´ıa et al. 2009). In
this episode, the sudden increase in the hard flux might be due
to the ejection of the inner hot disk or corona (which is often observed
in GX 339−4 during this state; Gallo et al. 2004) but the
radio observations during this episode were lacking and hence
an exact production mechanism for the second hard re-flare is
difficult to understand.

GX 339−4 showed A, B, and C types of QPOs during the
2006--2007 outburst (Motta et al. 2009). In three observations
where soft lags are observed, C-type QPOs are present (Motta
et al. 2009). In order to investigate the small dynamical changes
during the observations where lags are detected, we have
extracted the PDS in the initial (part A) and final (part B)
parts of the light curves. These fitted PDSs are plotted in
Figure 4. The PDSs are fitted with a broken power-law +
Lorentzian model.We found that the centroid QPO frequencies
are shifted to lower frequencies, implying that the physical
regions responsible for the QPOs have changed during the
respective observations (see Figure 4 and Table 1). Generally
it is observed that, as the QPO frequency increases, the width
of the QPO also increases, changing from a type-C QPO to
a type B and then to a type A one (Casella et al. 2005). The
QPO profile is often fitted with a Lorentzian but as its shape
changes from a type Cto type A, a Gaussian function is preferred
(Belloni 2010). In the second and third observations, the shift
in the QPO centroid frequency is relatively larger ($\delta$f$_{c}$ $\sim$ 1 Hz)
compared to the first observation ($\delta$f$_{c}$ $\sim$ 0.2 Hz; Figure 4).
The production mechanism of low-frequency QPOs (LFQPOs)
and the respective locations on the accretion disk are still
a matter of debate but often these QPOs are attributed to
the Comptonization region in the accretion disk, which is
generally assumed to be located at the inner region of the
disk (Chakrabarti \& Manickam 2000; Titarchuk \& Fiorito
2004). Sometimes, LFQPOs are related to the propagation of
a truncated disk front through the hot flow in the accretion
disk (Done et al. 2007).
During our analysis, we found that in two observations,
ObsId: 92085-01-01-05 and 92085-01-02-05, the QPO features
are present in the respective PDSs but are not mentioned in the
work of Motta et al. (2009). These observations were classified
as the HS state as per the Belloni (B) classification and the IM
state as per the McClintock and Remillard (MR) classification
scheme (see Table 1 in Motta et al. 2009). It is found that
in ObsID: 92085-01-02-05, the QPO feature is not present
in the 2.06--5.71 keV energy band but is seen only in the
6.12--14.76 keV band (single-bit mode data are used), whereas
in ObsID: 92085-01-01-05, the QPO feature is present in both
energy bands (generic binned mode data; Figure 5). The powerlaw
+ Lorentzian model is used to fit the PDS in ObsID: 92085-
01-01-05 and the broken power-law + Lorentzian model is used
in the other observation. In Figure 5, the respective PDSs are
fitted with and without the QPO feature. The dashed line shows
the fit with the QPO feature and the solid line shows the fit
without it. The residuals are plotted with respect to the simple
power-law and broken power-law models for each observation
(Figure 5). We find a $\delta$$\chi^{2}$ value of 82 ($\chi^{2}$ = 198.10 without
QPO, $\chi^{2}$ = 116.10 with QPO for dof 71) and $\delta$$\chi^{2}$ = 71.32
($\chi^{2}$ = 144.40 without QPO, $\chi^{2}$ = 73.08 with QPO for dof
63), respectively, for the two observations. The high $\delta$$\chi^{2}$ values
clearly indicate the presence of a QPO feature. The centroid
frequency of QPO in ObsID: 92085-01-01-05 is found to be
at f$_{c}$ = 4.91 $\pm$ 0.12 Hz with a quality factor Q = f$_{c}$/$\delta$f = 3.14, whereas in the other observation, the centroid frequency
is found to be at f$_{c}$ = 10.70 $\pm$ 0.24 Hz with a quality factor Q = f$_{c}$/$\delta$f = 2.20. Since the QPOs in the respective observations
are strong properties of the IM state, we conclude that these two observations belong to the IM state rather than the HS
state.
\begin{figure}
\centering
%\resizebox{\textwidth}{1}   
\includegraphics[height=15cm,width=10cm,angle=-90]{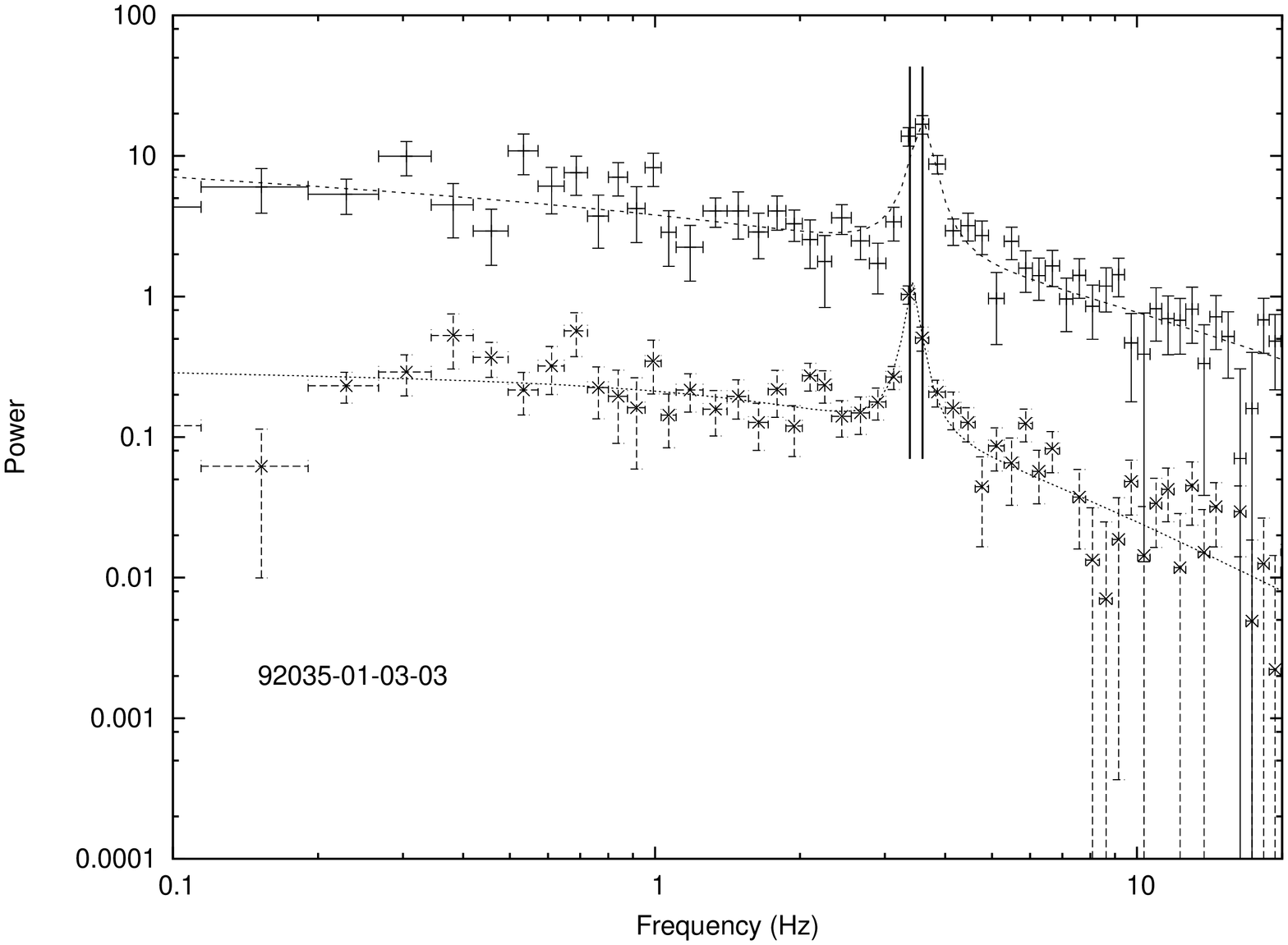}\\
\includegraphics[height=15cm,width=10cm,angle=-90]{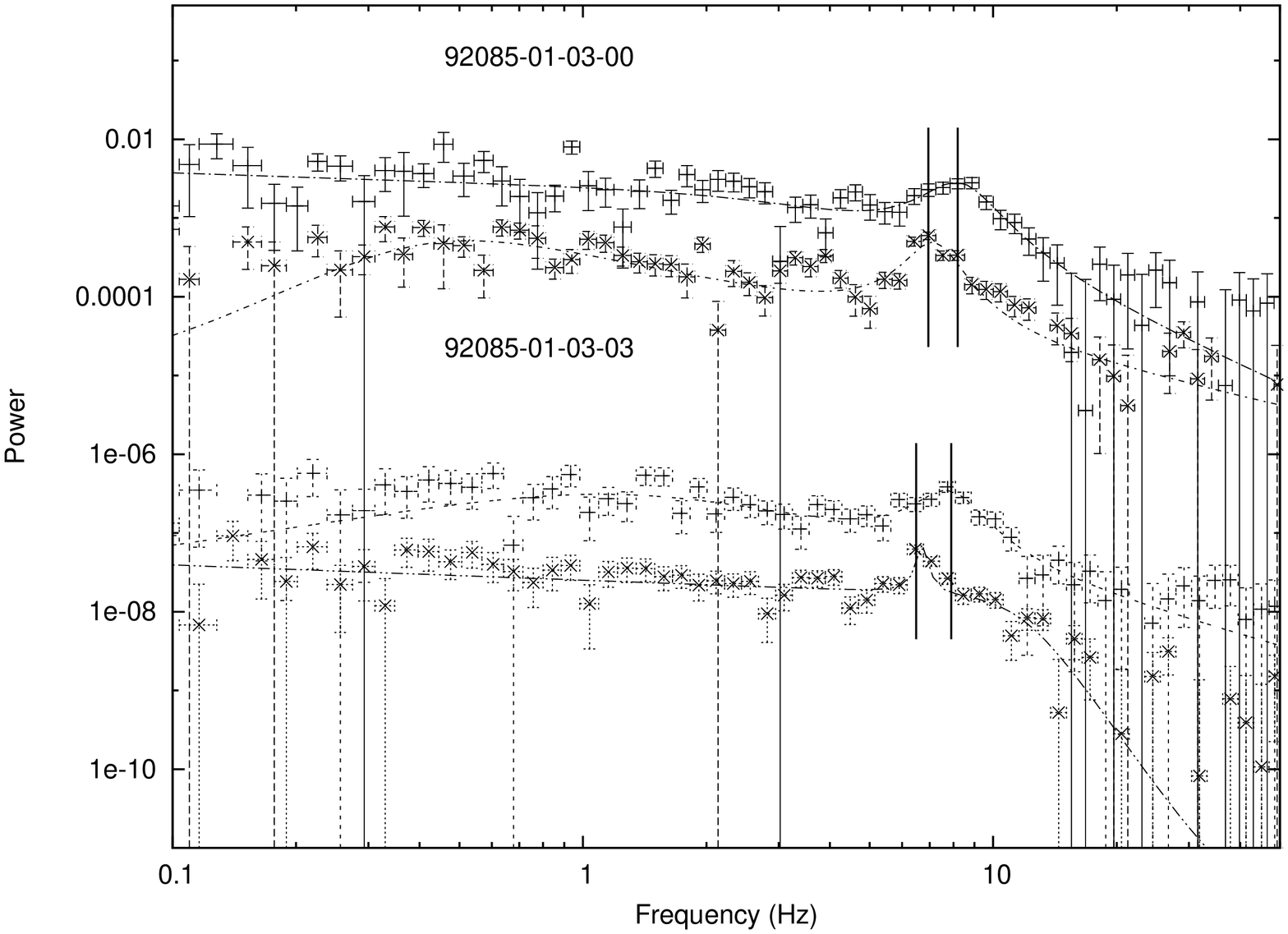}\\
     \caption{For the three observations, the power density spectrum are plotted 
(top panel :92035-01-03-03 and bottom panel:92085-01-03-00, 92085-01-03-03). 
For each observation, the upper PDS corresponds to part A and the lower PDS 
corresponds to part B sections of the light curves. The vertical line highlights
the shift in the observed centroid frequency. The power is  the 
normalized power in the units of
(rms/mean)$^{2}$ / Hz.
In the bottom panel, the PDS are appropriately shifted on the vertical scale for clarity. } 
       \label{Fig4}
 \end{figure}

\begin{figure}
\centering
%\resizebox{\textwidth}{1}   
\includegraphics[height=15cm,width=10cm,angle=-90]{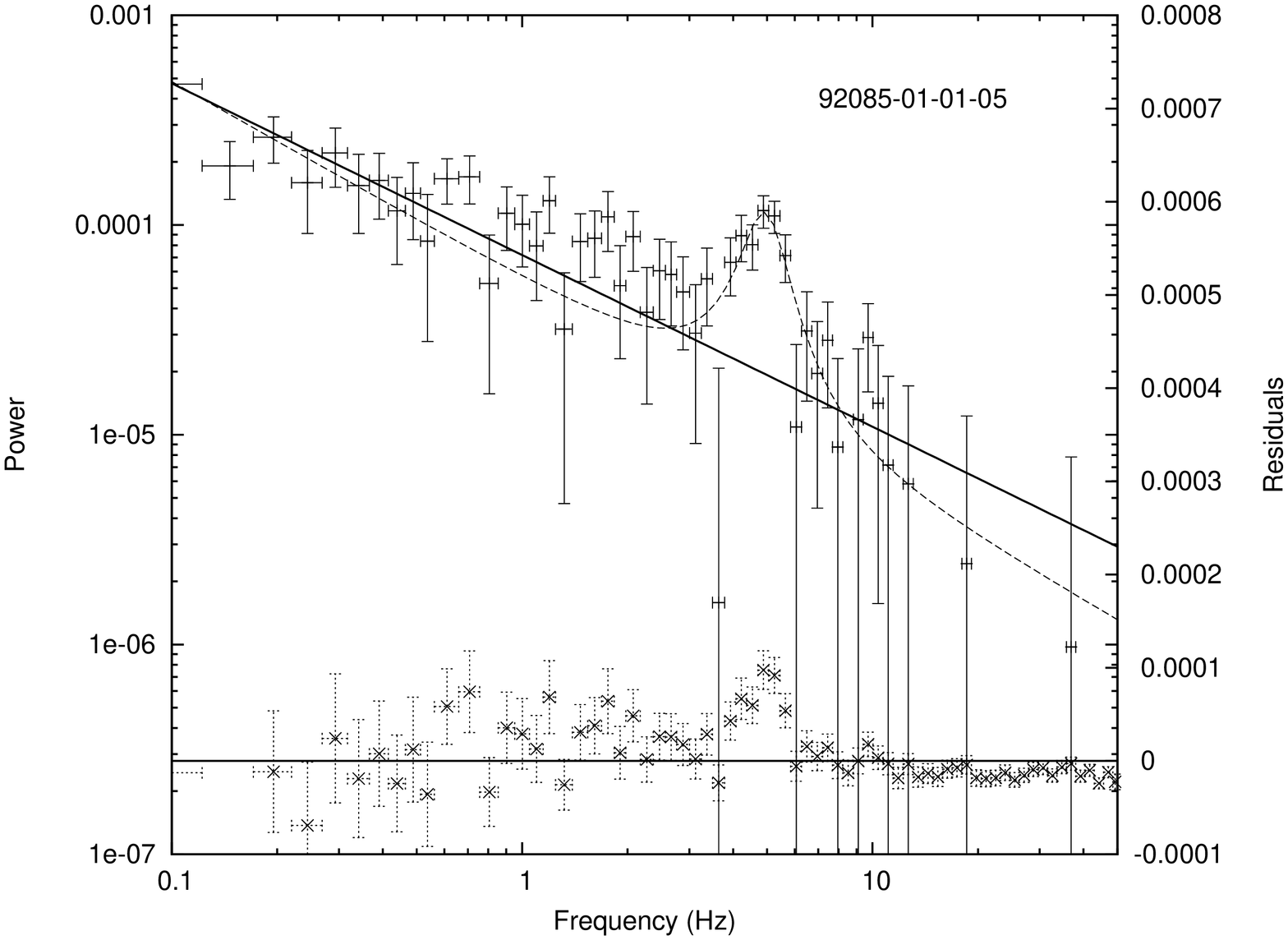}\\
\includegraphics[height=15cm,width=10cm,angle=-90]{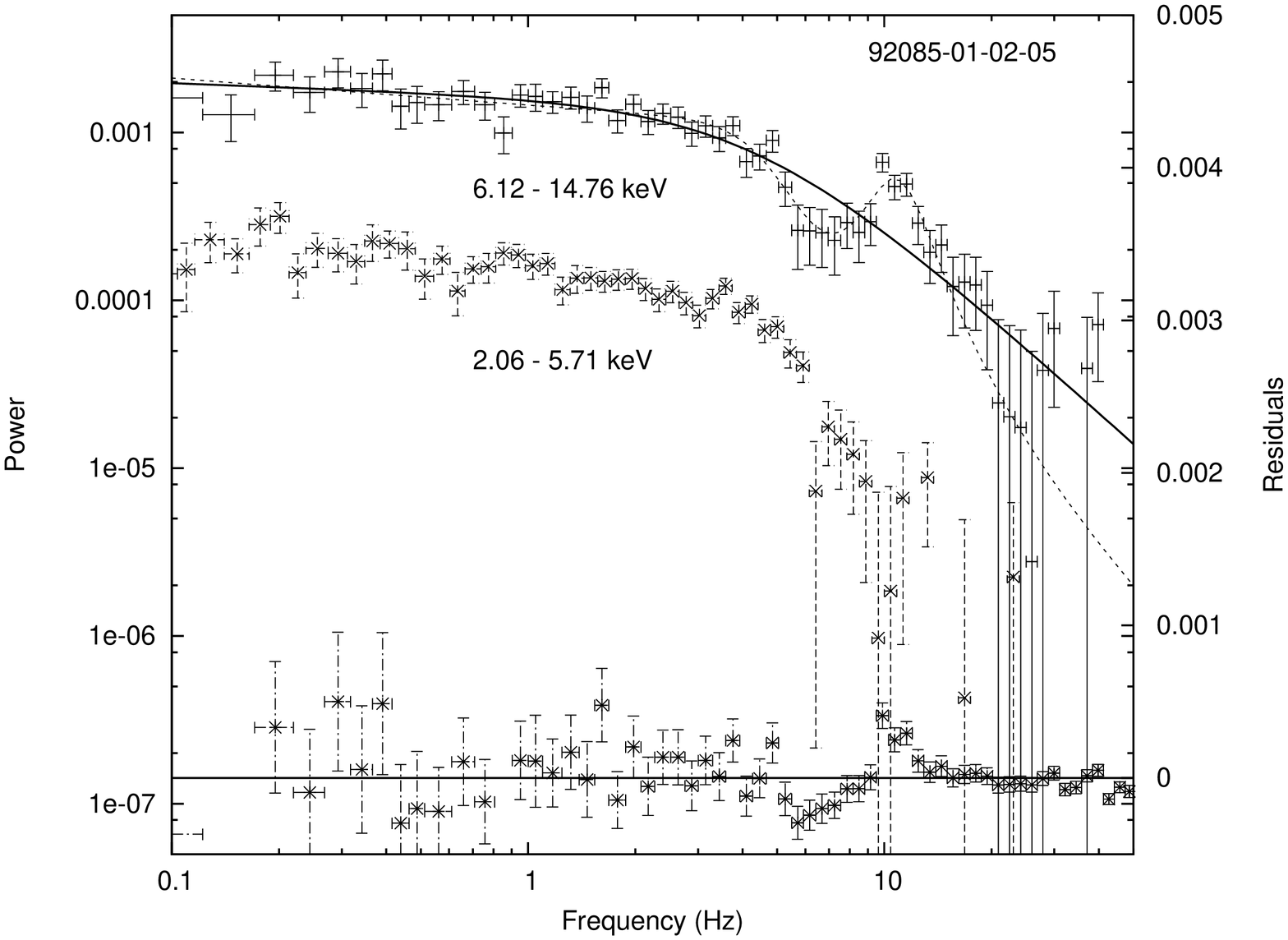}

     \caption{The figure show the fitted power density spectrum for two observations 
(top: 92085-01-01-05, bottom: 92085-01-02-05). For both the observations, the PDS is fitted with and without QPO feature. In each panel, the dash line 
shows the fit with QPO and the solid line shows the fit without QPO. 
In bottom panel, both the energy band PDS (2.014--5.71 keV PDS is shifted by a factor of 0.5 for clarity) 
are shown and QPO feature is clearly seen in higher energy band. 
 In the bottom of each panel, the residuals are clearly visible around the QPO feature. The Power is in units of (rms/mean)$^{2}$ / Hz.} 
       \label{Fig5}
 \end{figure}

\subsection{Spectral Analysis}

The temporal analysis shows large-scale lags and a prominent
shift in the centroid frequencies in the second and third
observations compared to the first. Hence, we perform a detailed
spectral analysis for the last two observations. The modelindependent
spectral ratios (the ratio of the part A spectrum to
the part B spectrum) are shown in Figure 6 and they show a
pivoting feature in the low energy range (4--5 keV) which indicates
that the probable changes are occurring at lower energies.
We have fitted the 2.8--30.0 keV spectra of the initial and final
parts (the same parts used to obtain the PDS) of the respective
observations and unfolded a model consisting of a disk blackbody,
power law, and absorption, along with a Gaussian line
(model: wabs*(diskbb+Gaussian+power-law in XSPEC). The
equivalent hydrogen column density is fixed at 0.5 $\times$ 10$^{22}$ cm$^{-2}$
(Dickey \& Lockman 1990), the Gaussian line energy is fixed
at 6.4 keV, and the remaining parameters were set free during
the fitting procedure. The unfolded spectra for parts A and
B are shown for the third observation (ObsID: 92085-01-03-
03) in Figure 7. It is clear from the figure that the pivoting
is caused by changes in the disk normalization and power-law
index components which explain the model-independent pivoting.
We found that all other parameters except for the disk
temperature varied significantly between parts A and B in the
individual observations (Table 1). The unabsorbed soft and hard
fluxes were anti-correlated between the initial and final parts
and it was found that the disk flux decreases with a simultaneous
increase in the power-law flux (Table 1). These differences
indicate that during the observed lag timescale the spectrum
has comprehensively changed. In order to show the prominent
differences in the spectra between parts A and B, we have used
the part A spectrum model parameters to fit the part B spectrum
and plotted the $\delta$ $\chi$ (i.e., residuals in terms of sigmas with
error bars) for both parts in Figure 8. It can be seen that there
is an excess in the 6--15 keV energy region, which indicates that 
during the lag timescale there is an increase in flux in the
said energy domain.
\begin{figure}
\centering
%\resizebox{\textwidth}{1}   
\includegraphics[height=15cm,width=10cm,angle=-90]{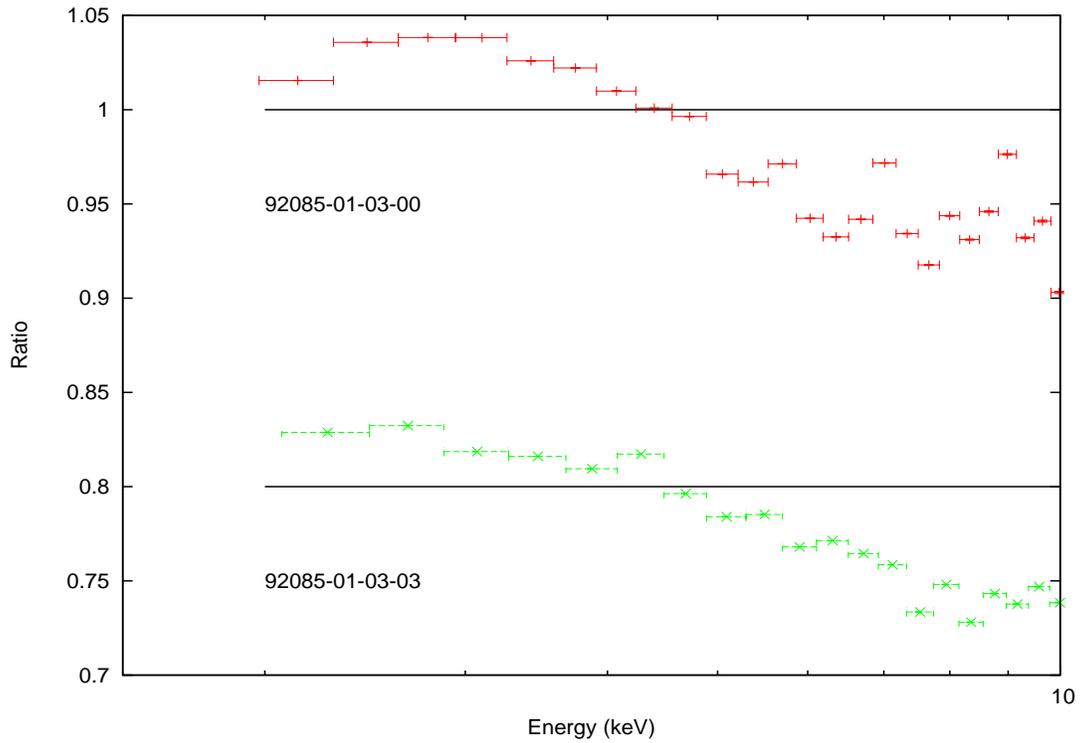}
     \caption{The ratios of the initial and the final parts (A \& B) spectra of 
the second and the third observations are shown. The data for the third 
observation are shifted vertically down for clarity. For both the observations 
the pivot point is found to be around 4-5 keV.} 
       \label{Fig6}
 \end{figure}

\begin{figure}
\centering
%\resizebox{\textwidth}{1}   
\includegraphics[height=15cm,width=10cm,angle=-90]{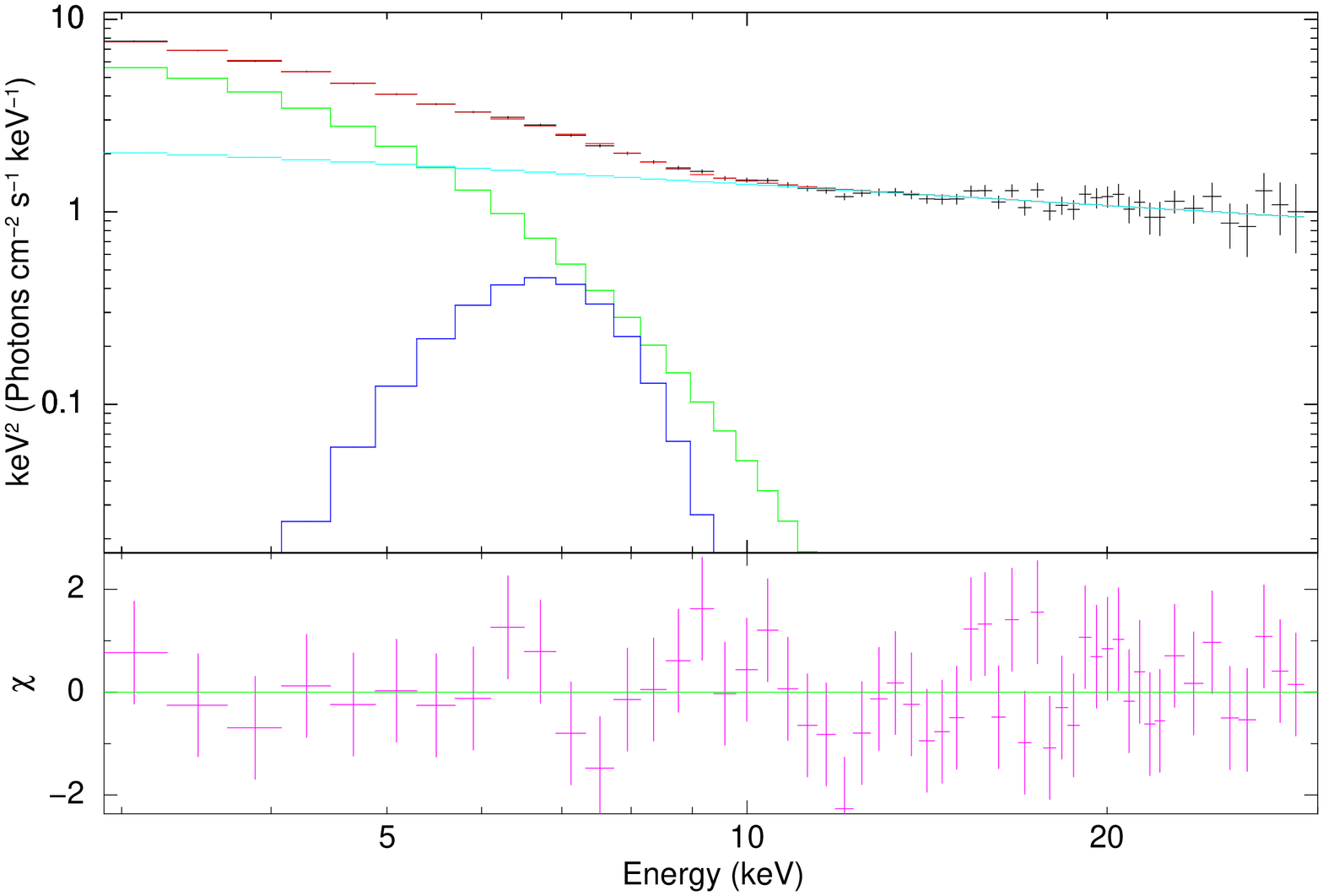}\\
\includegraphics[height=15cm,width=10cm,angle=-90]{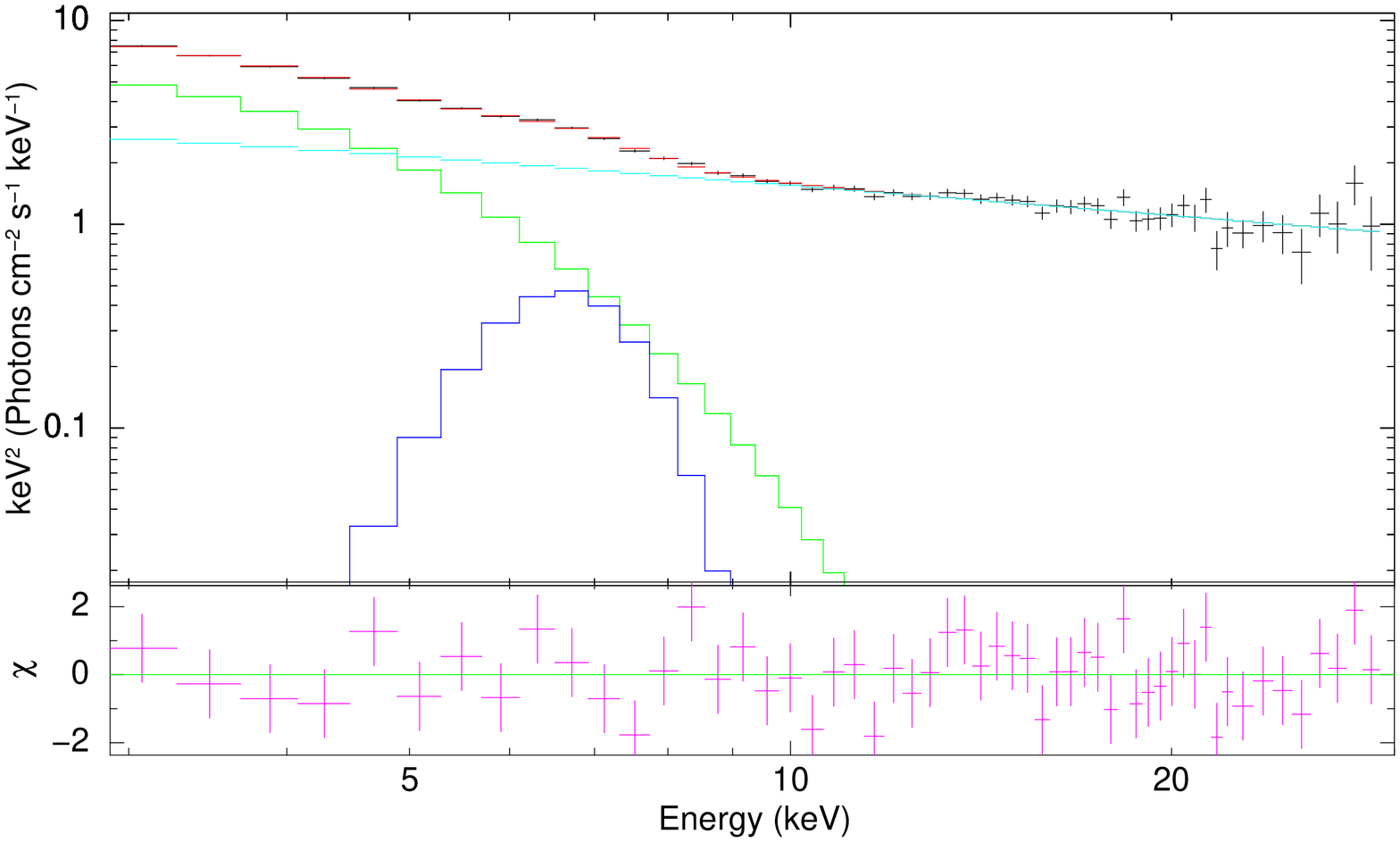}
     \caption{The unfolded spectra along with delta $\chi$ are shown for part A and B for the observation, ObsId: 92085-01-03-03. 
The model dependent pivoting point agrees with the model independent pivoting point in the respective spectra around $\sim$ 5 keV.} 
       \label{Fig7}
 \end{figure}

\section{Discussion and Conclusion}

The spectral and temporal properties of accretion disks in
black holes can be explained using the framework of a truncated
disk geometry. In the LH state, the disk is assumed to be
truncated at a large radial distance and as the disk approaches
the last stable orbit, it becomes non-truncated, resembling the
properties of the TD state. Recent RXTE, XMM-Newton, and
Chandra observations found a cool thermal component in the
LH state (Reis et al. 2009, and references therein) but more
observational studies are required to establish the geometrical
picture in this state. Our detailed spectral and temporal results
from the study of various black hole sources, GRS 1915+105,
XTE J1550−564, and H1743−322, show that in the IM state/
SPL state, the accretion disk is truncated very close to the
black hole and the truncation radius (r$_{t}$) is around $\sim$r$_{t}$ $\le$ 25r$_{s}$
(where r$_{s}$ is the Schwarzschild radius; Choudhury \& Rao 2004;
Choudhury et al. 2005; Sriram et al. 2007, 2009). Similar anticorrelated
hard as well as soft lags were found in one neutron
star source (Cyg X-2) during which the spectrum changes at
different pivoting energies (Lei et al. 2008). These results are
very important as they show that the accretion disk is truncated,
i.e., a radially moving Keplerian disk along with a Compton
cloud/base of jet most probably located close to the black hole.
In this paper, we report the discovery of anti-correlated soft
lags of the order of a fewhundred to thousand seconds (Figure 1).
The anti-correlated soft lag is further supported by the ISGRI/
INTEGRAL hard energy band in the third observation (ObsID:
92085-01-03-03, Figure 2). The CCF between the PCA/RXTE
(2--5 keV) and ISGRI/INTEGRAL (20--40 keV) was found to be
anti-correlated with a negative lag timescale. This lag timescale
is a few hundred seconds smaller than the lag obtained using
the PCA/RXTE data. This may be due to the fact that the
simultaneous observation times from RXTE and INTEGRAL
are of shorter duration. During the observed lags, the physical
properties of the inner region are changed; this fact is further
supported by the study of the PDS of the respective observations.
It was found that the QPO centroid frequencies have been
shifted to a lower frequency in a single observation. There is
no clear consensus as to which physical component (Compton
cloud or disk) is responsible for the production of LFQPOs.
Recently, Cabanac et al. (2010) showed that an oscillating
hot corona is able to produce the LFQPOs well, assuming a
truncated disk geometry. According to this model, a phase lag
of π between the low and high energy bands is seen (see their
Figure 8), which is a strong prediction of the truncated geometry.
These theoretically predicted variabilities closely match the first
observation light curves where a ∼100 s anti-correlated soft lag
was observed (Figure 1). In the Cabanac et al. (2010) model, the
QPO propagates toward higher frequency as the radius of the
corona front moves toward the black hole, whereas we find that a
shift in the disk inner radius is decreased ($\propto$ disk normalization)
for a few percentile change in the QPO frequency. This is due to
the fact that the boundary (inner front) where the disk truncates
and the Compton cloud dominates could be the same, and hence,
similar results were observed. The shift in the centroid frequency
to lower values (Figure 4) indicates that the corona front/disk
front ismoving away from the black hole, resulting in an increase
in the hard X-ray flux which is clearly seen in the $\delta$$\chi$ plot
(Figure 8).
In the spectral domain, it is clear that, except for the disk
temperature, all other parameters were significantly varied in
a single observation between parts A and B (Table 1). The
most important parameters are the soft and hard fluxes which
change in an opposite manner, i.e., the soft flux decreases with
an increase in the hard flux (Table 1). The power-law index
steepens with an increase in the power-law normalization. The
observed anti-correlated soft lags are associated with the spectral
variability during which the geometry of the accretion disk
changes. The gravitational heating and Compton cooling are
continuous processes in the accretion disk and the truncation
of the disk is most probably due to changes in the level of
Compton cooling mechanism, which is further related to the
disk soft photon flux (∝ mass accretion rate). As the disk size
increases, the number of soft photons increases, which cools
down the Compton cloud and vice versa. Our spectral results
show that the disk normalization decreased, implying that the
disk inner front is moving toward the black hole which cools
the Compton cloud (the power-law index steepens). The PDS
study shows that the QPO centroid frequency shifted to a lower
frequency which indicates an increase in the disk/corona inner
front. These observed contradictory features can be explained if
we assume that the inner hot flow (ADAF) condenses back onto
the inner disk, relatively increasing the size of the disk, steepens
the power-law index, and increases the hard flux. The IM state
is the most probable phase of accretion where the condensation
of the inner hot matter can transform into an inner disk
(Meyer-Hofmeister 2004; Meyer et al. 2007; Liu et al. 2007).
More studies are required to understand this phenomenon (if it
is true), especially in order to understand the properties of the
new disk (which formed from the condensation of hot matter)
and the indigenous disk (standard Keplerian disk).

\begin{figure}
\centering
%\resizebox{\textwidth}{1}   
\includegraphics[height=15cm,width=10cm,angle=-90]{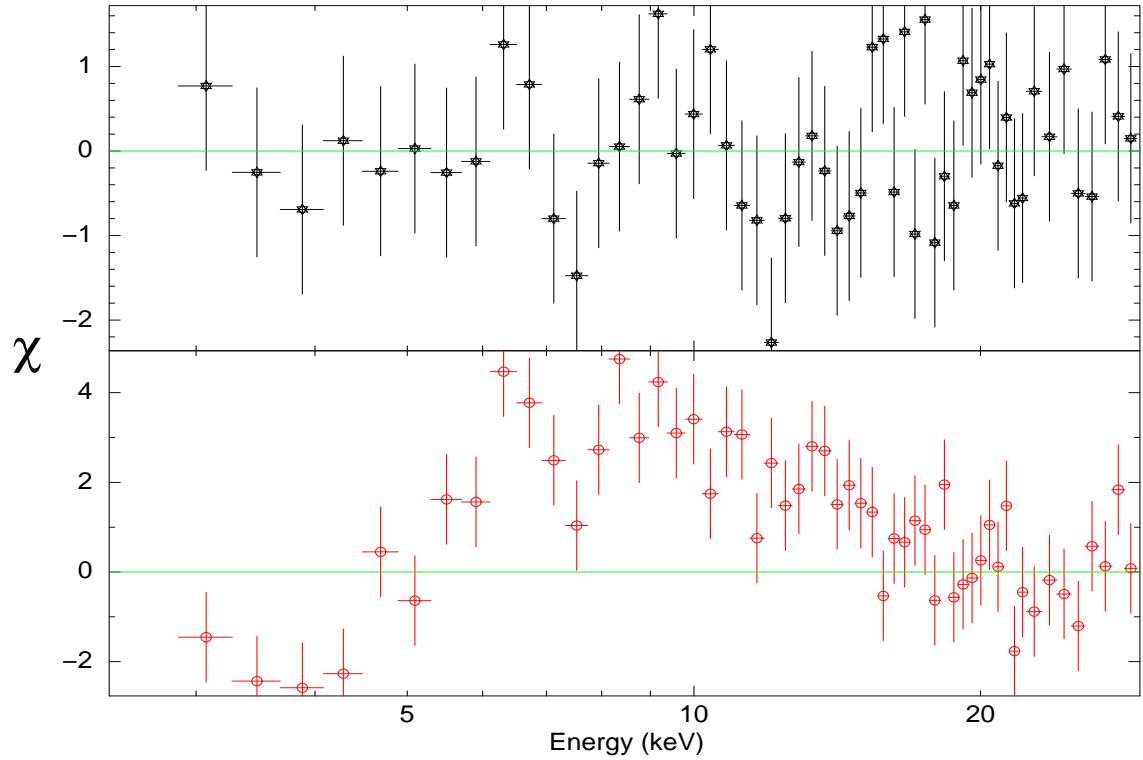}
     \caption{The figure shows the delta $\chi$ for part A (top) and part B 
(bottom) for the ObsId: 92085-01-03-03. In order to show the change in the part A and part B spectra, 
the part B spectrum is fitted using part A model parameter values. 
Clear excess is observed in part B in the energy range 6 -- 15 keV. 
} 
       \label{Fig8}
 \end{figure}

The observed lag timescales suggest that the geometry of
the accretion disk changes, which is strongly supported from
spectral studies. The lag was found when the source was in the
IM state. Our temporal and spectral variation during the anticorrelated
soft lag indicates a truncated accretion disk geometry
in the IM state. Similar kinds of studies can be extended to
neutron star systems and cataclysmic variables to constrain the
geometry of the accretion disk in a specific phase of accretion.

\acknowledgements 
We thank the anonymous referee for constructive and critical
comments. We acknowledge the use of RXTE data from the
HEASARC public archive. This work is also based on observations
made with INTEGRAL, an ESA science mission with
instruments and science data center funded by ESA member
states and with the participation of Russia and the USA. This
research has made use of data obtained through HEASARC
Online Service, provided by NASA/GSFC, in support of the
NASA High Energy Astrophysics Programs.

\setcounter{table}{0}
{\begin{table*}
\begin{minipage}[t]{\columnwidth}
\caption{Details of the spectral and temporal parameters in individual part of the respective observations. A and B corresponds to the initial and final parts of the observation.} 
\label{tab1}
\centering
\renewcommand{\footnoterule}{}
\begin{tabular}{ccccc}
\hline
Parameters&\multicolumn{2}{c}{92085-01-03-00} &\multicolumn{2}{c}{92085-01-03-03}\\
\hline
&A&B&A&B\\
\hline
$kT_{in}\footnote{Disk temperature using diskbb model}$&$0.87\pm0.007$ &$0.84\pm0.008$ &$ 0.86\pm0.006$ &$0.85\pm0.01 $\\
$N_{bb}$&$2014\pm100$ & $1827\pm111$&$ 1955\pm77$&$1758\pm83$ \\
%E$_{w}$ &$1.17\pm 0.21$ & 0.78$\pm$0.22 &1.04$\pm$0.17 & 0.82$\pm$0.16  \\
%$\Gamma_{th}\footnote{thcomp index}$&$2.53\pm0.05$&$ 2.54\pm0.04$&$ 2.07\pm0.01$&$2.05\pm0.01$&$2.12\pm0.01$&$2.10\pm0.01$\\
%$kT_{e}\footnote{Electron Temperature}$&$7.93\pm1.67$&$11.47\pm4.0$&$8.23\pm0.33 $&$6.94\pm0.28$&$9.44\pm0.91$&$8.95\pm0.90$\\
$\Gamma_{Pl}\footnote{Power-law index}$&2.42$\pm$ 0.05&2.59$\pm$0.04	 &2.40$\pm $0.04& 2.51$\pm$0.04\\
%$N_{th}\footnote{thcomp normalization}$&$3.77\pm1.50$&$11.5\pm4.1$&$ 0.74\pm0.01$&$0.75\pm0.26$&$1.04\pm0.20$&$0.85\pm0.24$\\

$N_{Pl}\footnote{Power-law normalization}$&$4.06\pm0.72$&6.46$\pm$0.80&3.56$\pm$0.42&5.08$\pm$1.02\\
$\chi^{2}$/dof&54/55 &40/55 &49/46 & 64/49\\
disk flux\footnote{The flux unit for all the models is $10^{-9}$ergs $cm^{-2} s^{-1}$}&9.15&7.34 & 8.78& 7.44 \\
%thcomp flux&41.17&46.50 & 10.11& 8.65 & 12.67 & 11.30&-&-\\
Power-law flux &8.47 &10.18 &7.95 & 9.26\\
\hline
%Simultaneous fit\\
%$N_{th}$&$11.52\pm3.5$&$10.85\pm3.0$&$0.84\pm0.04$&$0.47\pm0.05$&$1.07\pm0.22$&$0.94\pm0.22$&&\\
%$N_{bb}$ &$0.24\pm.03$ & $0.27\pm.04$ &$ 504\pm31$ &$444\pm27$ \\
%$kT_{in}$&$1.11\pm0.02$&$1.07\pm0.02$&$0.30$(fix)&$0.30$(fix)&0.28(fix)&0.28(fix)&&\\
%$\Delta$$N_{th}$/$N_{th}$(\%)&6.0&-&44&-&12&-&&\\
%$\Delta$$N_{bb}$/$N_{bb}$(\%)& & & &\\
%$\Delta$$kT_{in}$/$kT_{in}$(\%)&3.60&-&-&-&-&-&&\\
\hline 
Delay (sec)&$-1068\pm62$&-& $-858\pm108$ &-\\
f (Hz)\footnote{QPO centroid frequency, respective PDS are fitted with a power-law+Lorentzian model}&$8.15\pm0.33$&$7.04\pm0.15$ &$ 7.64\pm0.35$ &$ 6.67\pm0.18$ \\
%$\Delta$f/f\%& $-15.76$& --&$10.94$&-\\
\hline
\hline
\end{tabular}
\end{minipage}
\end{table*}}

%%%%%%%%%%%%%%%%%%%%%%%%%%%%
\end{document}